\documentclass[fleqn,usenatbib]{mnras}

\usepackage{newtxtext,newtxmath}

\usepackage[T1]{fontenc}
\usepackage{ae,aecompl}

\usepackage{graphicx}	
\usepackage{amsmath}	
\usepackage{amssymb}	

\title[PG\,1018--047: Evolutionary Constraints]{Evolutionary Constraints on the Long-period Subdwarf B Binary PG\,1018--047}

\author[Deca et al.]{J. Deca$^{1}$\thanks{E-mail: jandeca@gmail.com}, 
J. Vos$^{2}$, P. N\'emeth$^{3,4}$
P. F. L. Maxted$^{5}$, 
C. M. Copperwheat$^{6}$, 
\newauthor T. R. Marsh$^{7}$, and R. {\O}stensen$^{8}$\\
$^{1}$Laboratory for Atmospheric and Space Physics, University of Colorado, 3665 Discovery Drive, Boulder, CO  80303, USA.\\
$^{2}$Instituto de F\'{\i}sica y Astronom\'{\i}a, Universidad de Valparaiso, Gran Breta\~{n}a 1111, Playa Ancha, Valpara\'{\i}so 2360102, Chile.\\
$^{3}$Astroserver.org, 8533 Malomsok, Hungary.\\
$^{4}$Dr. Remeis-Sternwarte, Institute for Astronomy, University Erlangen-N\"urnberg, Sternwartstr. 7, 96049 Bamberg, Germany.\\
$^{5}$Astrophysics Group, Keele University, Keele, Staffordshire ST5 5BG, United Kingdom.\\
$^{6}$Astrophysics Research Institute, Liverpool John Moores University, IC2, Liverpool Science Park, 146 Brownlow Hill, Liverpool L3 5RF, UK.\\
$^{7}$Department of Physics, University of Warwick, Coventry CV4 7AL, United Kingdom.\\
$^{8}$Department of Physics, Astronomy, and Materials Science, Missouri State University, Springfield, MO 65804, USA.
}

\date{Accepted XXX. Received YYY; in original form ZZZ}

\pubyear{2017}

\begin{document}
\label{firstpage}
\pagerange{\pageref{firstpage}--\pageref{lastpage}}
\maketitle

\begin{abstract}
\noindent We have revisited the sdB+K-star long-period binary PG\,1018--047 based on 20 new high-resolution VLT/UVES spectra that provided regular coverage over a period of more than 26 months. We refine the period and establish that the orbit is significantly eccentric ($P\,=\,751.6\pm1.9\,\rm{d}$ and $e\,=\,0.049\pm0.008$). A simultaneous fit derived from the narrow metal lines visible in the spectrum of the sdB star and the metal lines in the red part of the spectrum that originate from the companion provides the mass ratio, $M_{\rm MS}/M_{\rm sdB} = 1.52 \pm 0.04$, for the system. From an NLTE model atmosphere analysis of the combined spectra, we find $T_{\rm eff}$\,=\,29900$\pm$330\,K, log\,$g$\,= 5.65$\pm$0.06\,dex and log($n_{\rm He}$/$n_{\rm H}$)\,=\,--3.98$\pm$0.16\,dex for the primary, consistent with a B-type hot subdwarf star. The spectral contribution of the companion is consistent with a K5V-type star. With the companion having a mass of only $\sim$0.7$\rm\,M_{\odot}$ this system lies close to the boundary below which stable Roche-lobe overflow (RLOF) cannot be supported. To model the evolution of such a system, we have extended earlier MESA models towards lower companion masses. We find that both phase-dependent mass loss during RLOF, when 30 to 40\% of the available mass is lost through the outer Lagrange point, and phase-dependent mass loss during RLOF in combination with a circumbinary disk of maximum $M_{\rm CB} = 0.001\rm \,M_{\odot}$ could have formed the PG\,1018--047 binary system.
\end{abstract}




\section{Introduction}

Subdwarf B stars (sdBs) are core-helium burning stars with thin hydrogen envelopes (typically $M_H < 0.02\rm\,M_{\odot}$) and masses close to the core-helium-flash mass $\rm\sim0.47\,M_{\odot}$~\citep{Saffer_etal_1994, Brassard_etal_2001}. They are situated on the so-called extreme or extended horizontal branch, between the main sequence (MS) and the white dwarf cooling track at the blue-ward extension of the horizontal branch. In contrast to spectra from Population I MS B stars with the same colour, the Balmer lines in sdB spectra are found to be abnormally broad due to high surface gravities ($\log g\simeq  5.0-6.0$)~\citep{Heber_etal_1984, Heber_1986, Saffer_etal_1994}. sdBs are observed in all galactic populations and are considered to be the main source for the UV-upturn in early-type galaxies~\citep{Green_etal_1986, Greggio_Renzini_1990, Brown_etal_1997}. For an in-depth review on hot sdBs, we refer the reader to~\citet{Heber_2009, Heber_2016, Geier_etal_2015} and references therein.\\

\noindent A large fraction of sdBs are found in binary systems~\citep{Maxted_etal_2001, Copperwheat_etal_2011}, making binarity a key ingredient to comprehend and explain their evolution. Current evidence suggests that all sdBs were formed through binary evolution or merging~\citep{Mengel_etal_1976, Paczynski_1976, Webbink_1984, Han_etal_2000, Han_etal_2002}. Even more so because single-star scenarios (see e.g.~\citealt{DCruz_etal_1996}) have only been able to provide ad hoc arguments as to how an sdB progenitor can ignite helium burning in its core on or near the giant branch and at the same time lose almost its entire hydrogen envelope. Using binary population synthesis models, \citet{Han_etal_2002, Han_etal_2003} define three channels leading to the formation of hot subdwarfs: the common envelope channel, leading to short-period (hours to days) close binaries with late-type main sequence companions or white dwarfs; the stable Roche-lobe overflow (RLOF) scenarios, producing long orbital periods (hundreds of days) and early-type main sequence or late-type giant companions (revised models by \citealt{Chen_etal_2013, Chen_etal_2014} including atmospheric RLOF show that periods up to 1600\,days are viable); and a He-core white dwarf merger scenario responsible for the observed population of single sdBs. Alternatively, also the $\gamma$-formalism~\citep{Nelemans_etal_2000, Nelemans_etal_2001, Nelemans_2010} and the hierarchical triple-star progenitor scenario~\citep{Clausen_Wade_2011} might help explain the observed population of sdB stars.\\

\noindent Focussing on the formation of long-period binaries, the theoretically predicted population does not seem to correlate well with the observed numbers, in contrast to short-period sdB binary estimates from binary population synthesis studies. \citet{Lisker_etal_2005} found that from 76 high-resolution VLT spectra of sdB stars only 24 showed the signature of an FGK companion, none of which showed any detectable RV variability. \citet{Han_etal_2003} predict that the majority of sdB binaries, between 60 and 70 per cent of the total, form via the first stable RLOF channel. To date, however, following the initial discovery of PG\,1018--047~\citep{Deca_etal_2012}, only a handful long-period systems have been characterized~\citep{Barlow_etal_2012, Barlow_etal_2013, Vos_etal_2012, Vos_etal_2013, Vos_etal_2014, Vos_etal_2017a}, compared to more than 130 short-period systems~\citep{Geier_etal_2011, Barlow_etal_2012}. Additionally, all current evolution models predict circular orbits, while ten out of the currently known eleven systems (including PG\,1018--047 discussed here) have significantly non-circular orbits (see Table~4 in~\citealt{Vos_etal_2017a} for an overview).\\

\noindent \citet{Vos_etal_2015} identify from the literature three proposed evolutionary mechanisms which could lead to eccentric long-period orbits. (1) A tidally-enhanced wind mass-loss mechanism, introduced by \citet{Soker_etal_2000}. Tidal forces in an eccentric system can increase the wind mass-loss before a contact binary is formed, and in response, a phase-dependent wind mass-loss then increases the eccentricity. The mechanism has been successfully applied to the He-WD binary IP~Eri to interpret its highly eccentric orbit~\citep{Merle_etal_2014, Siess_etal_2014}. (2) Also a varying (decreasing) mass-loss rate from the periastron to the apastron during the Roche lobe overflow phase might increase the orbital eccentricity~\citep{Bonacic_etal_2008}. (3) Third, the interaction of an sdB binary with a circumbinary (CB) disk is considered. Stable CB disks are commonly observed around post-AGB binaries (e.g.~\citealt{Hillen_etal_2014}). In this case, \citet{Vos_etal_2015} assume that disks were also present during the RGB evolution of the sdB progenitor and combine the eccentricity-pumping mechanisms explored for CB disks in post-AGB binaries~\citep{Dermine_etal_2013} with phase-dependent Roche lobe overflow. From their analysis using the binary module of the stellar evolution code MESA~\citep{Paxton_etal_2011, Paxton_etal_2013}, \citet{Vos_etal_2015} conclude that only the latter two, i.e. the models including the eccentricity-pumping processes  during and after RLOF, are able to form binary systems with an sdB primary and wide eccentric orbits.  \\

\section{Observations and data reduction}~\label{sec2}

\noindent PG\,1018--047, our target, was first discovered as an ultraviolet-excess stellar object in the Palomar-Green Survey \citep{GSL_1986}. The system has an apparent visual (Str\"omgren) magnitude $m_{y} = 13.32$. A first radial velocity analysis concluded that PG\,1018--047 might not have a short orbital period~\citep{Maxted_etal_2001}. The presence of weak spectral features from a late-type companion in the spectrum encouraged continued follow-up of the system to unravel the evolutionary mystery of the system. After a decade of monitoring with medium-resolution spectroscopy,~\citet{Deca_etal_2012} constrained the orbit of PG\,1018--047 to be 760 $\pm$ 6 days. Making use of the Balmer lines of the subdwarf primary and the narrow absorption lines of the secondary present in the spectra, they derived the radial velocity amplitudes of both stars, and estimated the mass ratio $M_{\rm MS}/M_{\rm sdB} = 1.6 \pm 0.2$. From the combination of visual and infrared photometry, the spectral type of the companion star was determined to be a mid-K dwarf. Unfortunately, the data quality was not sufficient to establish also the eccentricity of the binary orbit, possibly the determining factor to unravel the evolutionary scenario of the system. \\

\begin{figure*}
\centering
\includegraphics[width=0.9\textwidth]{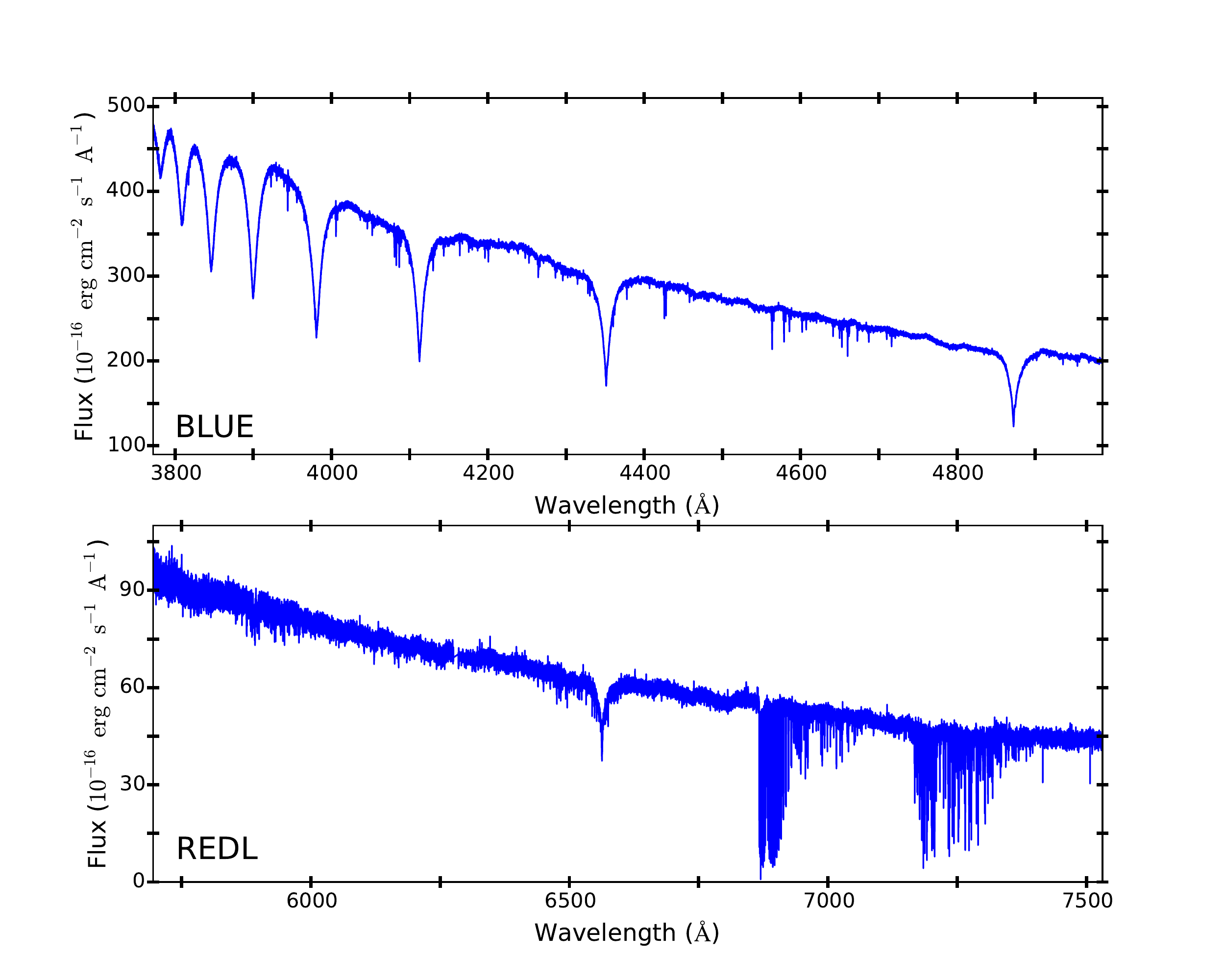}
\caption[]{Flux-calibrated sample spectrum of PG\,1018--047 obtained with VLT's UVES instrument.}
\label{fig:spectra}
\end{figure*}

\noindent PG\,1018--047 deserved the attention of high-resolution spectroscopy and using the Ultraviolet and Visual Echelle Spectrograph (UVES) mounted on the European Southern Observatory Very Large Telescope (VLT) 8.2-m telescope unit on Cerro Paranal/Chile~\citep{Dekker_etal_2000}, we obtained 20 spectra with each arm in service mode between December 21, 2010 and March 1, 2013. An overview is given in Table~\ref{tbl:dataoverview}. An exposure time of 1800 seconds was used for both the blue and red spectra. With a slit width of 0.8'' (corresponding to a full width at half maximum below 5\,km\,s$^{-1}$) under average observing conditions we obtained a signal-to-noise ratio of approximately 75 (90 in the blue), sufficient to obtain radial velocities good to better than 0.5\,km\,s$^{-1}$. The spectra ranged between $3760-4980$\,\AA\, in the blue arm and $5690-7530$\,\AA\, in the red arm (Figure~\ref{fig:spectra}). The reduction of the raw frames was conducted using the UVES Common Pipeline Library (CPL) recipes (version 4.1.0) within ESOReflex (version 2.6), the ESO Recipe Execution Tool. The standard recipes were used to optimally extract each spectrum~\citep{Larsen_etal_2012}. 

\begin{table*}
\caption[]{Overview of the VLT/UVES spectroscopic observations, including the radial velocity measurements for both the sdB primary and main sequence companion. Note, these radial velocity measurements are over an order of magnitude more precise than before (see Table~3 in~\citet{Deca_etal_2012}).}
\label{tbl:dataoverview}
\begin{tabular}{cccc}\hline\hline
Date			&	HJD - 245\,0000	&	RV$_{\rm MS}$ (km\,s$^{-1}$)	&	RV$_{\rm sdB}$ (km\,s$^{-1}$)	\\\hline
21/12/2010	&	5562.83253  		&	37.89 $\pm$ 0.69 			&	38.27 $\pm$ 0.22	\\
01/01/2011	&	5588.82119  		&	36.95 $\pm$ 0.45 			&	40.39 $\pm$ 0.11	\\
27/01/2011	&	5551.72459  		&	37.23 $\pm$ 0.99 			&	37.06 $\pm$ 0.30	\\
06/02/2011	&	5598.67253  		&	36.65 $\pm$ 0.81 			&	40.84 $\pm$ 0.20	\\
01/03/2011	&	5621.56080  		&	35.58 $\pm$ 1.79 			&	42.70 $\pm$ 0.23	\\
23/03/2011	&	5643.65802  		&	34.10 $\pm$ 0.96 			&	44.52 $\pm$ 0.31	\\
06/04/2011	&	5657.64707  		&	34.23 $\pm$ 1.31 			&	45.39 $\pm$ 0.15	\\
04/05/2011	&	5685.57795  		&	31.72 $\pm$ 0.92 			&	46.54 $\pm$ 0.28	\\
23/11/2011	&	5888.82064  		&	35.82 $\pm$ 1.03 			&	43.18 $\pm$ 0.14	\\
20/12/2011	&	5915.75611  		&	36.13 $\pm$ 0.71 			&	40.92 $\pm$ 0.22	\\
07/01/2012	&	5933.67543  		&	36.91 $\pm$ 1.13 			&	38.72 $\pm$ 0.15	\\
04/02/2012	&	5961.74056  		&	38.24 $\pm$ 1.07 			&	36.44 $\pm$ 0.07	\\
11/03/2012	&	5997.63130  		&	41.12 $\pm$ 0.77 			&	32.95 $\pm$ 0.15	\\
20/04/2012	&	6038.46524  		&	44.32 $\pm$ 0.86 			&	30.42 $\pm$ 0.14	\\
18/05/2012	&	6066.47992  		&	44.01 $\pm$ 0.78 			&	28.76 $\pm$ 0.09	\\
10/06/2012	&	6088.50008  		&	45.49 $\pm$ 0.82 			&	28.27 $\pm$ 0.11	\\
11/11/2012	&	6242.80334  		&	40.80 $\pm$ 1.35 			&	32.92 $\pm$ 0.13	\\
14/12/2012	&	6275.75715  		&	38.62 $\pm$ 1.07 			&	35.08 $\pm$ 0.16	\\
13/01/2013	&	6305.76808  		&	37.43 $\pm$ 0.63 			&	37.67 $\pm$ 0.17	\\
01/03/2013	&	6352.70847  		&	35.81 $\pm$ 0.69 			&	41.10 $\pm$ 0.19 \\\hline\hline
\end{tabular}
\end{table*}

\section{Observational analysis}~\label{sec3}

\subsection{Orbital solution}\label{sec:orbitalsolution}

The radial velocities (RVs) of both the sdB and the MS component of the system are determined following the procedures outlined in~\citet{Vos_etal_2012, Vos_etal_2013}. First, the reduced 1-D spectra are cleaned from cosmic rays and normalised by fitting and subtracting low order polynomials to obtain a straight baseline. Next the resulting spectra are cross-correlated with synthetic templates as the sdB component is primarily visible in the blue part of the spectrum, while the MS component contributes most to the red part of the spectrum. For the sdB component, we use the NLTE model derived in Section~\ref{sec:peter}, whereas we adopt a high-resolution spectrum with $\rm T_{eff} = 4700\,K$ and $\log g = 4.80\rm\,dex$ from the BlueRed library\footnote{The BlueRed library of high-resolution synthetic stellar spectra in the optical can be found at http://www.inaoep.mx/~modelos/bluered/documentation.html}~\citep{Bertone_etal_2008} for the companion.\\ 

\noindent For the sdB primary we adopt the following wavelength regions: $3910 - 3957\,$\AA, $3990 - 4082\,$\AA, $4149 - 4308\,$\AA, $4364 - 4477\,$\AA, $4584 - 4721\,$\AA. Note, these regions exclude the Balmer lines, historically used to obtain the sdB RVs (e.g.~\citealt{Deca_etal_2012}), but do include the characteristic helium I absorption lines as well as various oxygen, nitrogen and silicon lines. For the secondary RVs we use the region between $6095 - 6460\,$\AA. In addition, all interstellar lines within these regions are discarded as well. The final radial velocities are listed in the two rightmost columns of Table~\ref{tbl:dataoverview} for the sdB and MS companion, respectively.\\  

\begin{figure}
\centering
\includegraphics[width=\columnwidth]{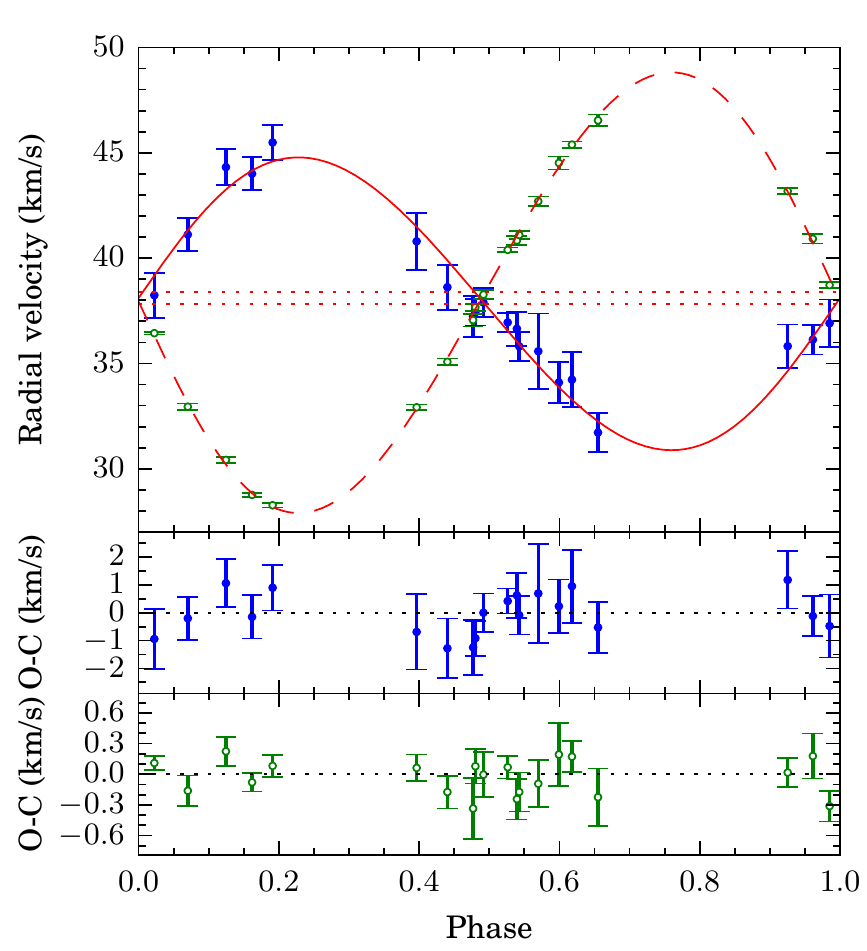}
\caption[]{The spectroscopic orbital solutions for PG\,1018--047. Top: the best fitting radial-velocity curves (solid line: MS, dashed line: sdB), and the observed radial velocities (blue filled circles: companion, green open triangles: sdB). The measured system velocity of the two components is shown by a dotted line. Middle: residuals of the MS component. Bottom: residuals of the sdB component.}
\label{fig:orbitalsolution}
\end{figure}

\noindent We calculate the orbital parameters of both binary components by fitting a Keplerian orbit to the RV measurements, simultaneously optimising the period ($P$), the eccentricity ($e$), the time and angle of periastron ($T_0$ and $\omega$), two amplitudes ($K_{\rm MS}$ and $K_{\rm sdB}$), and two systemic velocities ($\gamma_{\rm MS}$ and $\gamma_{\rm sdB}$). The orbital solution of~\citet{Deca_etal_2012} is used as a first guess for the fitting procedure. The RVs are weighted according to their errors following the relation $w = 1/\sigma$. The \citet{LucySweeney_1971} eccentricity test is used to verify whether the orbit is significantly eccentric. During fitting we permit the system velocities of both components to vary independently, in this way allowing for gravitational redshift effects~\citep{Vos_etal_2013}. The uncertainties on the final parameters are estimated using 2000 Monte-Carlo iterations on the perturbed radial velocity errors. We refer the reader to \citet{Vos_etal_2012, Vos_etal_2013} for more details on the procedure. Based on the best-fitting Kepler curve, we find $P_{\rm orb}=752\pm2\rm\,days$ with a significant ($p<0.0007$) eccentricity $e=\,$0.049$\,\pm0.008$. The average deviation between the observed velocities and the fit for one observation is $0.03\rm\,km\,s^{-1}$ for both the sdB and companion. The rms error of the MS component is $0.75\rm\,km\,s^{-1}$, the rms error of the sdB component is $0.17\rm\,km\,s^{-1}$.
The orbital solution of PG\,1018--047 is given in Table~\ref{tbl:orbitalsolution} and plotted in Figure~\ref{fig:orbitalsolution}. 

\begin{table}
\centering
\caption{Spectroscopic orbital solution for the sdB primary and the MS companion of PG\,1018--047.}\label{tbl:orbitalsolution}
\begin{tabular}{lrr}\hline\hline
	Parameter				&	sdB				&	MS				\\\hline
$P$ (d)					&	\multicolumn{2}{c}{751.6$\pm$1.9}		\\
$T_0$					&	\multicolumn{2}{c}{2455193$\pm$16}		\\
$e$						&	\multicolumn{2}{c}{0.049$\pm$0.008}			\\
$\omega\,(^{\circ})$				&	\multicolumn{2}{c}{92$\pm$8}			\\  
${\rm d}$               &   \multicolumn{2}{c}{590$\pm$30\,pc} \\
$q\,(M_{\rm MS}/M_{\rm sdB})$	&	\multicolumn{2}{c}{1.52$\pm$0.04}			\\
$\gamma\,\rm(km\,s^{-1})$	    &	38.38$\pm$0.06	&	37.82$\pm$0.19	\\
$K\,\rm(km\,s^{-1})$			&	10.46$\pm$0.09	&	6.95$\pm$0.40	\\
$a\,\sin\,i\,(\rm R_{\odot})$		&	155$\pm$1		&	103$\pm$6		\\
$M\,\sin^3\,i\,(\rm M_{\odot})$ 	&	0.16$\pm$0.02 	&	0.25$\pm$0.01	\\
$\chi^2$                        &   1.03            &   0.65            \\ 
$\alpha$\,(J2000)        &       \multicolumn{2}{c}{10h\,21m\,10.587s} \\
$\delta$\,(J2000)         &       \multicolumn{2}{c}{-04$^{\circ}$\,56'\,19.56''} \\
$\mu_{\alpha}$\,(mas\,yr$^{-1}$)  &       \multicolumn{2}{c}{-15.0$\pm$2.1} \\
$\mu_{\delta}$\,(mas\,yr$^{-1}$)   &       \multicolumn{2}{c}{-11.9$\pm$2.6}\\\hline\hline
\end{tabular}
\end{table}

\subsection{Model atmosphere analysis}\label{sec:peter}

The high-resolution UVES spectra are dominated by the hot subdwarf and many weak lines of the cool companion can be identified superimposed on the sdB spectrum. 
Therefore, given the high resolution and signal-to-noise ratio of the UVES spectra, the spectral decomposition task is relatively easy for PG\,1018--047. To fit the composite spectrum we applied \textsc{NLTE Tlusty/Synspec}~\citep{HubenyLanz_1995, LanzHubeny_2007} model atmospheres and synthetic spectra with the \textsc{XTgrid} $\chi^{2}$-minimization fit procedure described in~\citet{Nemeth_etal_2012}. The procedure fits the observations with successive adjustments of the model parameters along the steepest gradient in a multi-dimensional parameter space. Along with the \textsc{NLTE} model for the sdB star, \textsc{XTgrid} also includes pre-calculated synthetic spectra for the cool companion. From the two models a synthetic composite spectrum is created and iteratively fitted to the observations.\\

\begin{table}
\centering
\caption[]{List of the strongest lines in the theoretical sdB spectrum. Only lines with a higher equivalent width (W$_\lambda$) than 30 m\AA\ are listed. 
}\label{tbl:peterlines}
\begin{tabular}{clr clr}\hline\hline
$\rm\lambda$ & Element  & W$_\lambda$   & $\rm\lambda$ & Element & W$_\lambda$ \\
(\AA) &  & (m\AA)  &  (\AA) &  & (m\AA) \\\hline
4088.862    &   \ion{Si}{IV} & 49.1 &  5001.474    &   \ion{N}{II}  & 91.7 \\
4267.183    &   \ion{C}{II}  & 84.9 &  5482.292    &   \ion{Ni}{III} & 40.9 \\
4284.979    &   \ion{S}{III} & 49.8 &  5875.615    &   \ion{He}{I}   & 108.6  \\
4552.622    &   \ion{Si}{III}& 85.7 &  5889.951    &   \ion{Na}{I}   & 65.6 \\
4649.135    &   \ion{O}{II}  & 95.5 &  6562.813    &   \ion{H}{I}    & 2574.3 \\
\\\hline\hline
\end{tabular}
\end{table}

\begin{figure*}
\centering
\includegraphics[width=\textwidth]{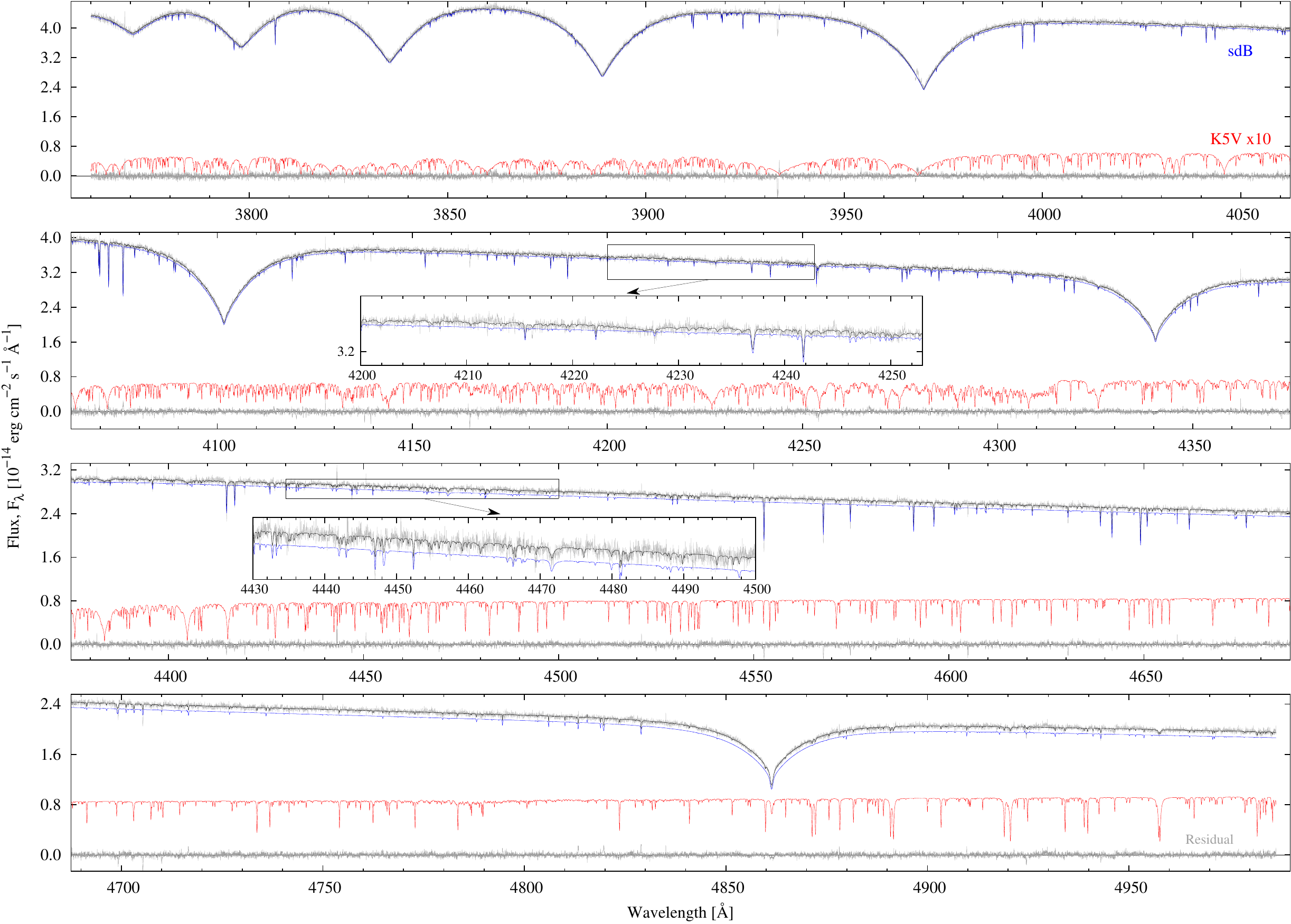}
\caption[]{The radial-velocity corrected UVES spectrum of PG\,1018--047 (grey). The subdwarf (blue) dominates the spectrum in the entire range while the K5V companion (red) contributes only $4.7 \pm 0.5\%$ of the total flux at 4980\,\AA. The thin black line overplotted on the observation is the best fit \textsc{Tlusty/XTgrid} binary model, the sum of the two components. To make the companion distinguishable from the residual its flux is multiplied by ten. The peak SNR of the UVES spectrum reaches $\sim$105 in the red. The atmospheric parameters of the binary members are given in Table~\ref{tbl:petertable2}.}
\label{fig:peter1}
\end{figure*}

\noindent \textsc{Tlusty} calculates plane-parallel metal line blanketed \textsc{NLTE} model atmospheres in hydrostatic and radiative equilibrium that are appropriate for hot stars. Our models include opacities from the 17 astrophysically important elements in the model atmosphere calculations and the first 30 elements in the spectral synthesis. The sharp metal absorption lines of the sdB star indicate a low rotational velocity, therefore we assume a non-rotating primary. The fit procedure starts with an initial model and iteratively updates the parameters along the steepest-gradient of the global chi-squared. The \textsc{BlueRed} grid allows us to fit the temperature, gravity and scaled-solar metallicity of the cool companion along with the \textsc{Tlusty} model for the primary. The fit is based on spectral lines only, because a reliable blaze correction of the UVES echelle spectra is not available. 
The strongest lines in the sdB model are listed in Table~\ref{tbl:peterlines}.
The best fit is shown in Figure~\ref{fig:peter1} and the final parameters are listed in Table~\ref{tbl:petertable2}. For a better representation of the observed data the continuum of the combined UVES observations is adjusted to the composite model in Figure~\ref{fig:peter1}.\\

\noindent We update \textsc{XTgrid} here and add a method to treat the different radial velocities of the binary members in multiple spectra as the included observations cover about two years, comparable to the orbital period. We utilize the high resolution synthetic spectral library (\textsc{BlueRed},~\citealt{Bertone_etal_2008}). These models cover late-type stars and were calculated with \textsc{Atlas9/Synthe}~\citep{Kurucz_1993}. Next, we fit the BLUE arm spectra (See Table~\ref{tbl:dataoverview}) simultaneously and measure the surface temperature and gravity of both members and their flux contribution to the composite spectrum. We determine individual abundances for the sdB star and estimate the metallicity of the cool companion. A more detailed description of the XTgrid code is given in \citet{Vos_etal_2017b}.\\

\noindent The low contribution of the cool companion to the BLUE arm spectra puts limits on our parameter determination. Even a relatively low, $\sim$20 km\,s$^{-1}$ projected rotation can change the derived metallicity and surface gravity significantly. To cope with this degeneracy we need observational data and synthetic spectra reaching to the near infrared, which are beyond the scope of this paper and the coverage of the \textsc{BlueRed} library.\\

\begin{table}
\centering
\caption[]{Parameters for the fits shown in Figure~\ref{fig:peter1}, with respect to the solar abundances from \citet{Asplund_etal_2009}.}\label{tbl:petertable2}
\begin{tabular}{lrrrrl}\hline\hline
Parameter Value		&	Value	&	+1$\sigma$	&	-1$\sigma$	&	Unit	&	$\times$ Solar		\\\hline
sdB:					&			&				&				&		&					\\
$T_{\rm eff}$			&	29870	&	300			&	360			&	K	&					\\
log g					&	5.68		&	0.03			&	0.09			&	dex	&					\\
log n(He)/n(H)			&	-3.98		&	0.160		&	0.160		&	dex	&	0.001			\\
log n(C)/n(H)			&	$<$ -5.3	&	-			&	-			&	dex	&	$<$ 0.02			\\
log n(N)/n(H) 			&	-4.71		&	0.038		&	0.038		&	dex	&	0.286			\\
log n(O)/n(H) 			&	-4.25		&	0.025		&	0.025		&	dex	&	0.116			\\
log n(Ne)/n(H) 			&	-4.97		&	0.420		&	0.420		&	dex	&	0.125			\\
log n(Si)/n(H) 			&	-5.35		&	0.067		&	0.067		&	dex	&	0.137			\\
log n(S)/n(H) 			&	-5.83		&	0.069		&	0.069		&	dex	&	0.111			\\
log n(Fe)/n(H) 			&	$<$ -5.6	&	-			&	-			&	dex	&	$<$ 0.08			\\
log n(Ni)/n(H)			&	$<$ -4.9	&	-			&	-			&	dex	&	$<$ 8.06			\\\hline
K5V:				&			&				&				&		&					\\
$T_{\rm eff}$			&	4240		&	250			&	250			&	K	&					\\
log g 				&	4.9		&	0.2			&	0.4			&	dex	&					\\
$\rm [Fe/H]$				&	$\leq$ -1		&				&				&	dex	&					\\
$F_{\rm sdB}/F_{\rm MS}$ (5000\,\AA)	&	20.3 	&	0.1	&	0.1			&		&					\\\hline\hline
\end{tabular}
\end{table}

\noindent The fit procedure converges on a typical sdB, but with a very low helium abundance, as expected from its unusually weak He\,{\sc i} lines. The spectral decomposition provides a K5V type main sequence companion. Using the mass ratio obtained in Section~\ref{sec:orbitalsolution} and assuming ${\rm M}_{\rm MS}=0.7$ M$_\odot$\footnote{Note, our modeling procedure does not allow an accurate error estimate. Adopting the estimates from~\citet{Habets_Heintze_1981} ($\pm0.3$\,M$_\odot$ for our spectral type) would lead to a $\pm0.2$\,M$_\odot$ error on the sdB mass.} for the K5V secondary we find $M_{\rm sdB}=(0.7\pm0.3)\,\times\,(0.66\pm0.04) = 0.46\pm0.03\,$M$_\odot$, quite close to the canonical sdB mass~\citep{Han_etal_2002}. Since we have $M \sin^3 i$ available we can obtain a rough estimate for the inclination of the system. From Table~\ref{tbl:orbitalsolution}:

\begin{align*}
\sin^3 i &= M/M_{\rm sdB} =(0.16\pm0.02)\,/\,(0.46\pm0.03) = 0.35\pm0.05\\
\sin^3 i &= M/M_{\rm MS} =(0.25\pm0.01)\,/\,0.7 \qquad\quad\quad\, = 0.36\pm0.02.
\end{align*}

\noindent Hence averaging $\sin i=0.71\pm0.02$ we find an inclination $i=45^{\circ}$.\\

\noindent \citet{Deca_etal_2012} found a correlation between optical, infrared colour indices and the flux contribution of cool main sequence Population I companions. They derived a K3--K6 type main-sequence companion from colour indices and estimated the flux contribution of the secondary to be $6.1\pm1.0$ \% in the $V$ band. We find here a K5V type template with $4.7\pm0.5$ \% contribution at 4980\,\AA,
which is fully consistent with the photometric constraints. \\

\subsection{Spectroscopic distance}

\noindent We calculate the spectroscopic distance to PG\,1018--047 by a direct scaling of the synthetic composite spectrum to the flux calibrated {\sc UVES} observations. The spectral decomposition allows us to measure the flux ratio ($f_r$) from the composite spectrum:

\[
f_r=\frac{F_{\rm MS}{R_{\rm MS}}^2}{F_{\rm sdB}{R_{\rm sdB}}^2}
\]

\noindent where $F$ denotes the flux and $R$ the stellar radius. We used the mass ($M_{\rm sdB}=0.46\pm0.03\,$M$_\odot$) and surface gravity ($\log{g}_{\rm sdB}=5.68\rm\,dex\,(cgs)$) of the primary to derive its radius ($R_{\rm sdB}=0.161\,$R$_\odot$). With this subdwarf radius and the surface flux from the subdwarf model, the total flux from the companion ($F_{\rm MS}$) is calculated. From the ratio of the observed flux ($f$) and the composite model flux, then, the scale factor ($F_{\rm sdB}\,(1 + f_r)\,f^{-1}$) and the distance are obtained:

\[
d = 
R_{\rm sdB}\sqrt{ \frac{F_{\rm sdB}(1 + f_r)}{f} }
\]

\noindent The dust maps of~\citet{Schlegel_etal_1998} provide an extinction $E({\rm B-V})=0.05\pm0.034$\,mag toward PG1018-047, which we also consider in our distance measurement. We calculate the scale factor around 4850\,\AA\, where the effects of reddening, atmospheric extinction and spectral lines are the lowest in the observation. The different radii of the subdwarf and the cool companion are implicitly included in the composite model and set up by the fit procedure. We find a scale factor of $(2.0\pm0.3)\times10^{21}$. With the subdwarf radius calculated above, the spectroscopic distance becomes ${\rm d}=590\pm30$\,pc. 
The formal error reflects the uncertainty in the scaling, but not in the assumed subdwarf radius and the flux calibration, which dominates the systematic errors. Note that the extinction numbers of \citet{Schlegel_etal_1998} are calculated towards infinity. Hence, at ${\rm d}=590\pm30$\,pc there may very well be somewhat less dust between Earth and our target, PG\,1018--047, and $E({\rm B-V})=0.05\pm0.034$\,mag should be taken as an upper limit to the extinction. Recalculating the spectroscopic distance without taking into account the extinction, we get ${\rm d}=633\pm 35$\,pc.

\subsection{Spectral energy distribution}

\noindent The photometric spectral-energy distribution (SED) of PG\,1018-047 can be used to determine the effective temperature and surface gravity of both components. This way an independent set of spectral parameters can be obtained to confirm the results of the \textsc{XTgrid} analysis. The photometry is collected from the literature using the subdwarf database\footnote{http://catserver.ing.iac.es/sddb/} \citep{Oestensen2006} supplemented with photometry obtained from the AAVSO Photometric All-Sky Survey (APASS) DR9 \citep{Henden2016}. All available photometry for PG\,1018-047 is summarised in Table\,\ref{tb:photometry}.\\

\begin{table}
\caption{Photometry of PG\,1018-047 collected from  APASS, 2MASS and str\"{o}mgren photometry from sdB candidates in the Palomar-Green survey \citep{Wesemael1992}.}
\label{tb:photometry}
\centering
\begin{tabular}{lrrr}
\hline\hline
\noalign{\smallskip}
Band    &   Magnitude    &   Rms error   \\
        &  mag      &   mag \\\hline
\noalign{\smallskip}
APASS $\rm B$       &  13.143   &   0.022  \\ 
APASS $\rm V$       &  13.311   &   0.036  \\ 
APASS $\rm g'$       &  13.084   &   0.016  \\ 
APASS $\rm r'$       &  13.483   &   0.017  \\ 
APASS $\rm i'$       &  13.718   &   0.044  \\  
STROMGREN $u$   &  13.161   &   0.006  \\ 
STROMGREN $b$   &  13.234   &   0.005  \\ 
STROMGREN $v$   &  13.224   &   0.010  \\ 
STROMGREN $y$   &  13.320   &   0.004  \\
2MASS $\rm J$      &  13.298   &   0.026  \\ 
2MASS $\rm H$      &  12.980   &   0.027  \\ 
2MASS $\rm K_s$      &  12.928   &   0.033  \\
\hline
\end{tabular}
\end{table}

\noindent To fit the observed photometry with a synthetic SED calculated from model atmospheres the same procedure as described in \citet{Vos_etal_2012, Vos_etal_2013} is used. Kurucz atmosphere models \citep{Kurucz1979} are used for the MS component while TMAP (T\"{u}bingen NLTE Model-Atmosphere Package, \citealt{Werner2003}) atmosphere models are used for the sdB component. The MS models have a temperature range of 3000 to 9000 K, and range in surface gravity from $\log{g}$=2.0 dex (cgs) to 5.0 dex (cgs). For the sdB models the ranges are T$_{\rm eff}$ = 20\,000 to 50\,000 K and $\log{g}$ = 5.0 to 6.5 dex (cgs). To derive the fluxes from the observed magnitudes, the filter profiles and zero points from the filter-profile service of the virtual observatory\footnote{http://svo2.cab.inta-csic.es/theory/fps/} \citep{SVO_Filter_Profile_Service} 
are used. \\

\noindent The SED fit includes eight parameters, the effective temperatures ($T_{\rm{eff,MS}}$ and $T_{\rm{eff,sdB}}$), surface gravities ($g_{\rm{MS}}$ and $g_{\rm{sdB}}$) and radii ($R_{\rm{MS}}$ and $R_{\rm{sdB}}$) of both components, the interstellar reddening $E(B-V)$ and the distance ($d$) to the system. As the mass ratio of the system is known, it is used to link the radii of the components to their surface gravity, reducing the number of free parameters to six.\\

\noindent The uncertainties on the parameters are calculated using 2D confidence intervals. The cumulative density function is used to calculate the probability of a model to obtain a certain $\chi^2$ value. The resulting cumulative probability distribution for each parameter pair is then used to derive 3\,$\sigma$ confidence intervals.\\ 

\begin{figure*}
\centering
\includegraphics{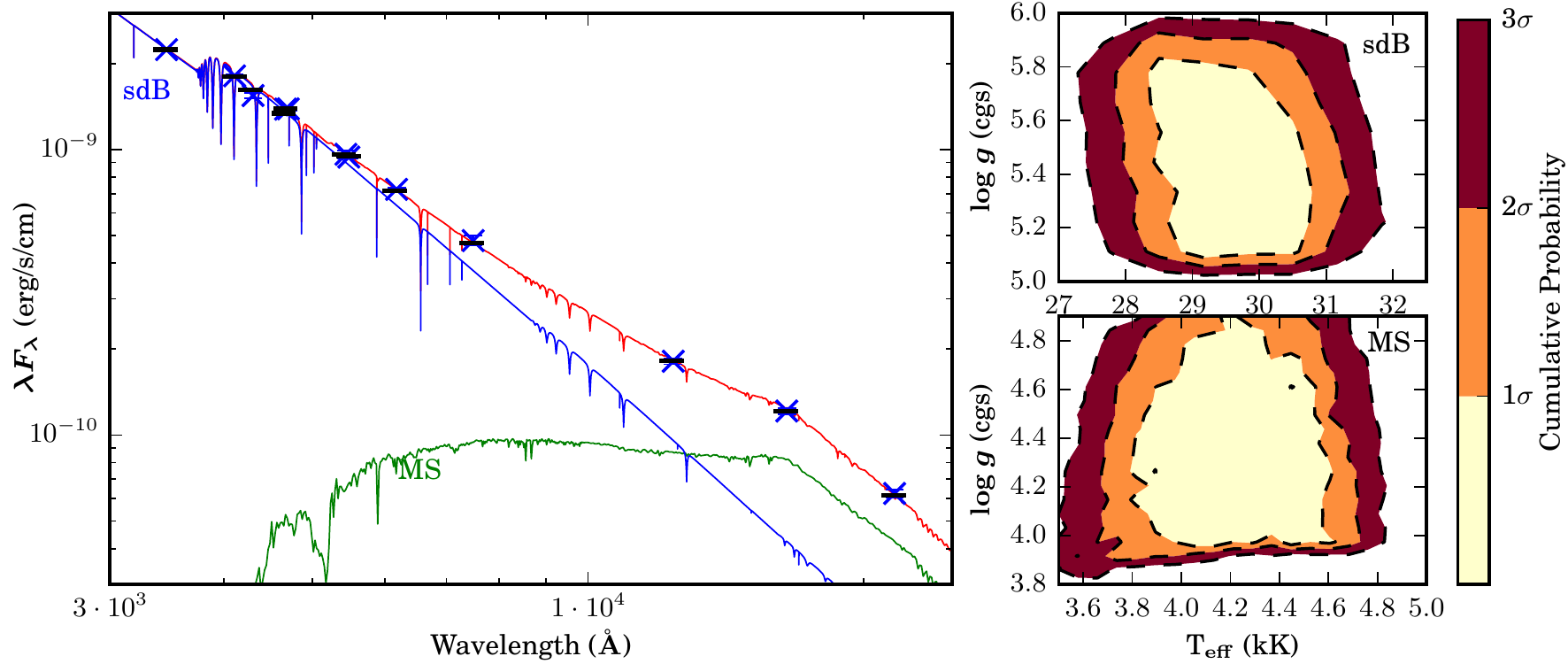}
\caption[]{The photometric SED of PG\,1018-047 with the best fitting model (left panel) and the confidence intervals (CI) for the effective temperature and surface gravity of both components (sdB: right top panel, MS: right bottom panel). The observed photometry is plotted in blue crosses, while the synthetic best fit photometry is plotted in black horizontal line. The best fitting binary model is shown in a red line, while the models for the cool companion and sdB star are show respectively in green line and blue line. In the CI plots, the cumulative probability of an area containing the correct solution is shown for the 1, 2 and 3 sigma confidence levels.}
\label{fig:sed-fit}
\end{figure*}

\noindent  The resulting spectral parameters for the sdB component of PG\,1018-047 are T$_{\rm eff}$ = 29500 $\pm$ 3000 K and $\log{g}$ = 5.70 $\pm$ 0.40, while for the MS component we find T$_{\rm eff}$ = 4150 $\pm$ 500 K and $\log{g}$ = 4.60 $\pm$ 0.50 dex (cgs). The best fitting model and the cumulative confidence intervals for the effective temperature and surface gravity of each component are shown in Figure~\ref{fig:sed-fit}. The spectral parameters derived from the photometric SED are in very good agreement with those derived with \textsc{XTgrid} based on the UVES spectroscopy. As both methods are independent, this provides a high confidence for these results.

\section{MESA modelling analysis}

\noindent The low mass of the companion and the small but non-zero eccentricity of PG\,1018-047 make this an interesting system to explore theoretical models. The larger the mass difference between the sdB progenitor and its companion, the higher the likelihood of unstable mass transferred~\citep{Ge_etal_2010,Ge_etal_2015}. Systems such as PG\,1018-047 are in that sense good test cases for the stability of mass transfer during RLOF. Compared to the models presented in \citet{Vos_etal_2015} this system fits in the low-period range, and therefore falls in the region covered by the RLOF models. These models, however, were calculated for systems with a companion mass of 0.8\,M$_{\odot}$ or higher.  Here we recalculate the models assuming a companion mass of 0.7\,M$_{\odot}$, consistent with the mid-K spectral classification derived above.\\

\noindent \citet{Vos_etal_2015} showed that two physical processes are capable of explaining the eccentricity of long-period post-RLOF systems: phase-dependent mass loss during RLOF on its own, or in combination with a CB disk that interacts with the binary. The short-period systems with low eccentricities, however, could only be formed by the phase-dependent mass-loss process \citep[See Figure 9 in][]{Vos_etal_2015}. We examine here these two mechanisms for the case of PG\,1018-047, taking into account also the lower companion mass of the binary. We extend the parameter space for the CB-disk models to include the short-period and low-eccentricity range of the system. Note, similar to the original article, the CB disk is always combined with phase-dependent mass loss as the disk is formed by the mass lost during RLOF. \\

\noindent Both phase-dependent RLOF and the interaction between a CB disk and the binary need a minimum eccentricity to work. We impose therefore $e_{\rm min} = 0.001$ for all models, i.e., if the eccentricity of the model drops below $e = 0.001$ it is considered circular. We elaborate on the impact of the latter in Section~\ref{sec:CB_disks}. Due to numerical constraints, we adopt an upper limit of $\dot{M}_{\rm max} = 10^{-2}\rm\,M_{\odot}\rm\,yr^{-1}$ on the mass-loss rate during RLOF as well. Finally, models with a Roche-lobe overflow radius of $R_{\rm donor} / R_{\rm RLOF} > 1.25$ are considered to be unstable, and are rejected. These criteria are identical to \citet{Vos_etal_2015}. \\

\subsection{Phase-dependent RLOF}

\noindent Most important for the phase-dependent RLOF process are the mass-loss fractions. During RLOF mass can be accreted by the companion or lost from the binary system altogether. For the latter we use the formalism of \citet{Tauris_vandenHeuvel_2006}, which describes three channels of mass loss to infinity.  Mass is lost either from the direct neighbourhood of the companion star ($\alpha$), from around the donor star ($\beta$) or through the outer Lagrange point ($\delta$). The difference between these three channels is the specific angular momentum lost from the system, which depends on the mass ratio ($q$) as follows:

\begin{align*}
\dot{J}_{\alpha} &\sim \frac{1}{1 + q}\\
\dot{J}_{\beta}  &\sim \frac{q^2}{1 + q}\\
\dot{J}_{\delta} &\sim 1+q
\end{align*}
For a given mass ratio, mass lost through the outer Lagrange point ($\delta$) will remove significantly more angular momentum from the binary than mass lost from around the donor ($\alpha$).

\begin{figure}
\centering
\includegraphics[width=\columnwidth]{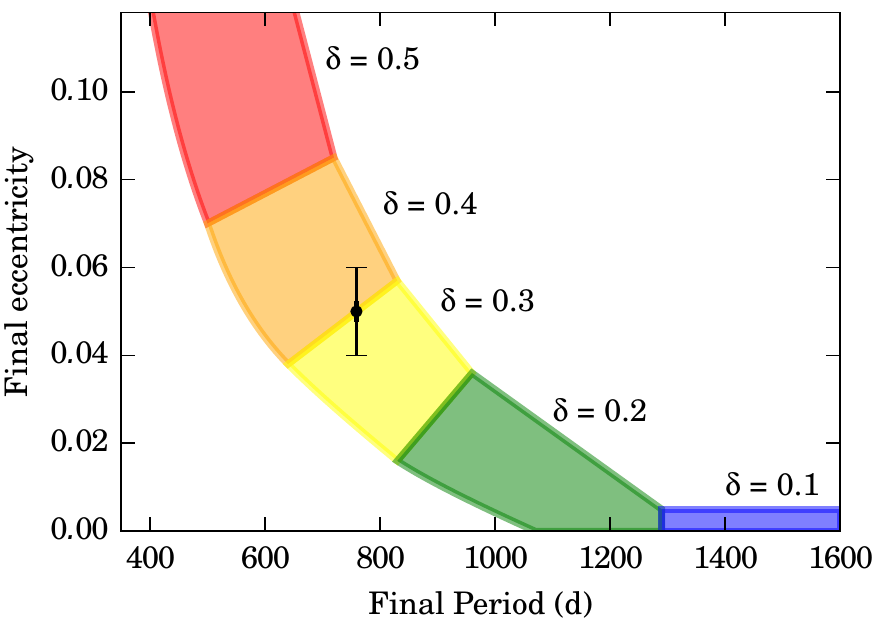}
\caption[]{The region of the period-eccentricity diagram that can be covered by the models with only phase-dependent RLOF. The regions where models with different values for mass-loss fraction $\delta$ are located are shown in different colours. These are not hard boundaries, but indicate where the majority of the models have that value for $\delta$. The period and eccentricity of PG\,1018--047 are plotted in black.}
\label{fig-mesa_rlof_delta}
\end{figure}

\noindent For this analysis a set of 300 models is calculated with different input parameters for the ranges given in Table\,\ref{tbl:orbitalsolution}. For the RLOF models, the disk parameters are fixed and the disk mass is set to 0. The models concur that PG\,1018-047 can form only if a large amount of mass is lost through the outer Lagrange point (fraction $\delta$) . In Figure~\ref{fig-mesa_rlof_delta} the ranges in period and eccentricity of models with different values for mass-loss fraction $\delta$ are indicated in different colours. The borders of these ranges are not exact, but aim to indicate how $\delta$ influences the final period and eccentricity of the system. We find that PG\,1018--047 may exist only if 30 to 40\% of the available mass is lost during RLOF and through the outer Lagrange point. \\

\noindent The other mass-loss fractions cannot be constrained for PG\,1018--047, as changing one can be counteracted by changing one or several of the other parameters. The amount of angular momentum lost from the donor or the accretor region is much smaller than when the mass leaves the system through the outer Lagrange point. The effect of changing $\alpha$ or $\beta$ on the final period and eccentricity, hence, is much smaller. \\

\subsection{Circumbinary disks}\label{sec:CB_disks}

\begin{figure}
\centering
\includegraphics{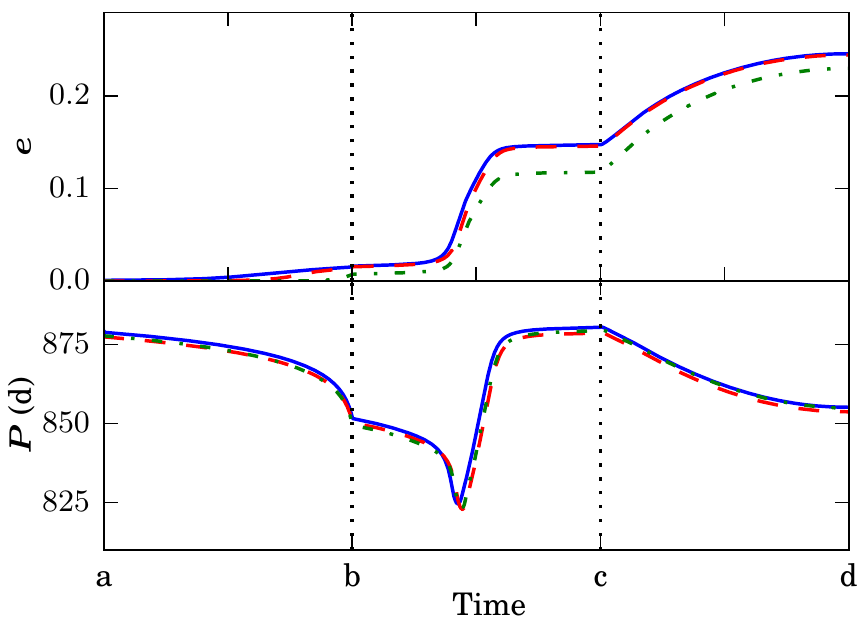}
\caption[]{Time evolution of the eccentricity (top panel) and orbital period (bottom panel) during the life time of the CB disk for a model with $e_{\rm min} = 10^{-3}$ (blue full line), a model with $e_{\rm min} = 10^{-4}$ (red dashed line), and a model with $e_{\rm min} = 10^{-5}$ (green dot-dashed line). Four different events are indicated on the x-axes. 
  a: $\dot{e}_{\mathrm{disk}} > \dot{e}_{\mathrm{tidal}}$, 
  b: $\dot{e}_{\mathrm{ml}} > \dot{e}_{\mathrm{tidal}}$, 
  c: $\dot{e}_{\mathrm{ml}} < \dot{e}_{\mathrm{tidal}}$ and 
  d: $\dot{e}_{\mathrm{disk}} < \dot{e}_{\mathrm{tidal}}$. 
  The time scale differs between the phases, but is linear within each phase. See also \citet{Vos_etal_2015} for an extensive description of the model.}
\label{fig-mesa_min_ecc}
\end{figure}

\noindent In the models presented by \citet{Vos_etal_2015}, PG\,1018-047 falls outside of the region that could be covered by the CB-disk models. The addition of a CB  disk to the phase-dependent RLOF models, however, is interesting as it almost completely removes the model dependence on the minimum eccentricity assumed in the pure phase-dependent RLOF models above. We demonstrate the effect in Figure~\ref{fig-mesa_min_ecc}, where the standard model from \citet{Vos_etal_2015} is plotted using three different values for the minimal eccentricity: $e_{\rm min}$ = 10$^{-5}$, 10$^{-4}$ and 10$^{-3}$. The evolution is shown when eccentricity pumping is active. After point d the eccentricity is quasi-steady over time, as the tidal forces are too weak to still significantly change the eccentricity. The final eccentricity and orbital period for the three models shown differs only by 0.01 in eccentricity and 5 days in orbital period. The weak dependency on the minimum eccentricity originates partly in the eccentricity dependence of the eccentricity pumping effect of the CB disk. The pumping efficiency rises steeply at low eccentricities, and peaks at $e = 0.03$ \citep[see Figure 1 of][]{Dermine_etal_2013}. For larger eccentricities its efficiency decreases. Models with a higher minimal eccentricity will reach this tip quicker, after which the disk contribution to the eccentricity pumping process diminishes. Changing the minimum eccentricity also alters the mass-loss rates during RLOF, the eccentricity pumping due to phase-dependent mass loss, and the tidal forces. Combining all these effects, however, the different models reach similar final periods and eccentricities.\\

\begin{figure}
\centering
\includegraphics{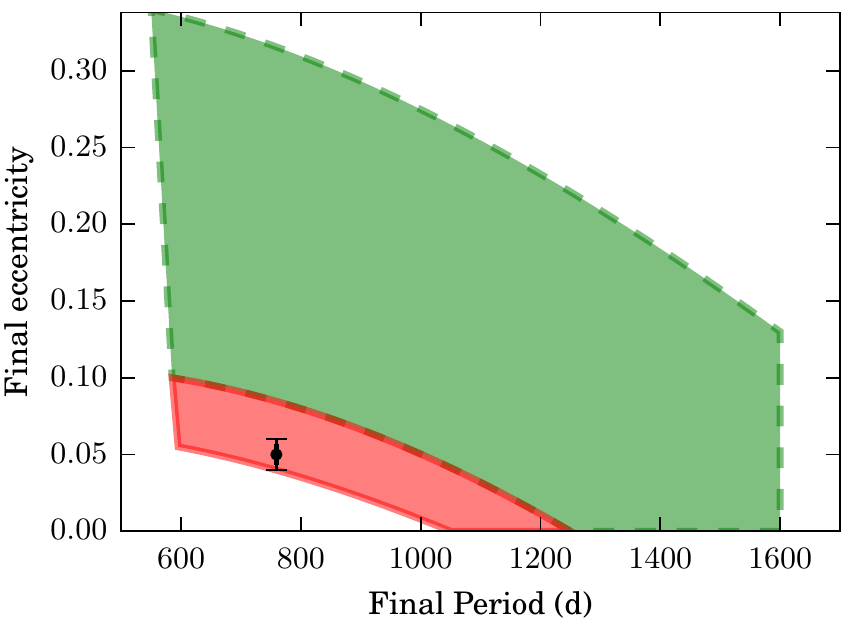}
\caption[]{The region of the period-eccentricity diagram that can be covered by the CB-disk models. The original models from \citet{Vos_etal_2015} are shown by the green area with dashed border, while the new models with lower CB disk masses are located in the red area with solid border. The location of PG\,1018-047 is plotted in black.}
\label{fig-mesa_disk_cbt_cbm}
\end{figure}

\noindent To reach the short period and low eccentricity of PG\,1018-047 we extend the parameter ranges from \citet{Vos_etal_2015}. We adapt two parameters that are currently badly constrained in literature. (1) Shortening the lifetime of the CB disk decreases the timespan over which the eccentricity pumping is effective, thus lowering the final eccentricity of the system as well. Stopping eccentricity pumping when the eccentricity reaches PG\,1018--047 value of $e = 0.05$ implies that the lifetime of the disk has to be shorter than 1000 years. This is too short to be considered a real CB disk, and does not correspond with observations of CB disks observed for post-AGB binaries~\citep{Gielen_etal_2008}. (2) Alternatively we can reduce the total mass in the CB disk. In previous models we assumed a CB disk mass of 0.01 $\rm M_{\odot}$, but direct measurements of CB disk masses allow rather large uncertainties still. Lowering the total disk mass to 0.001\,$\rm M_{\odot}$, forming a binary system with the properties of PG\,1018--047 through the CB-disk channel seems feasible. In addition, we find an upper limit for the total mass of PG\,1018--047's CB disk equal to $M_{CB} = 0.001\rm \,M_{\odot}$. In Figure~\ref{fig-mesa_disk_cbt_cbm} the region of the period-eccentricity diagram that can be covered by CB disk models with masses ranging from 0.0005 to 0.005 $\rm M_{\odot}$ is shown in red, while the original range of the models from \citet{Vos_etal_2015} is shown in green.\\

\section{Discussion}~\label{sec4}

\noindent Having constrained the orbital parameters of PG\,1018--047 as a non-circular 752-day binary system with low, but significant eccentricity, this indicates viable formation mechanisms. Many evolutionary mysteries, however, remain. \\ 

\subsection{A halo or disk star?}

\noindent One of these mysteries is why PG\,1018--047 appears to have such a low Helium abundance.
Could it be that the binary is a member of the Galactic halo population rather than a thick/thin disk system? \citet{Hog_etal_2000} list the proper motion of PG\,1018--047 as
\begin{eqnarray*}
(\mu_{\alpha},\mu_{\delta}) 	&=& (-15.1,-11.9)\pm(2.1,2.6)\,\rm{mas\,yr^{-1}}\\
						&=& (-38,-30)\pm(6,7)\,\rm{km\,s^{-1}}.
\end{eqnarray*}

\noindent Assuming $\gamma=38.2\pm0.2\,\rm{km\,s^{-1}}$ and a spectroscopic distance $\rm d = 590\pm30\,pc$ (Table~\ref{tbl:orbitalsolution}), these numbers yield a galactic space velocity~\citep{Johnson_Soderblom_1987}:

\begin{equation*}
(U,V,W) = (-26,-55,-11)\pm( 6, 5, 5) \,\rm{km\,s^{-1}},
\end{equation*}
\noindent with the U-axis defined towards the galactic centre. Converting to the local standard of rest~\citep{Dehnen_Binney_1998}, we obtain

\begin{equation*}
(U,V,W)_{\rm LSR} = (-16,-50, -4)\pm(6,5,5)\,\rm{km\,s^{-1}}.
\end{equation*}

\begin{figure}
\centering
\includegraphics[width=\columnwidth]{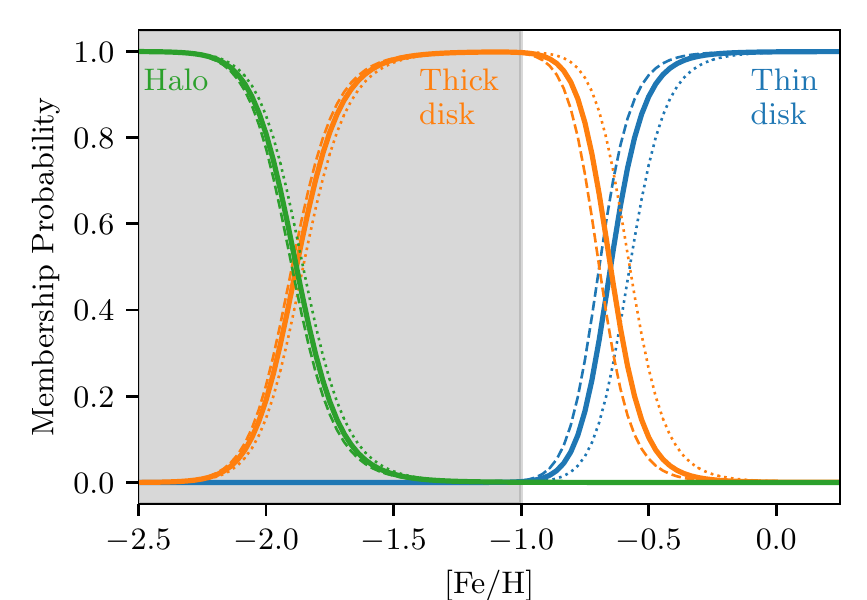}
\caption[]{The probability of PG\,1018-047 ([Fe/H] $\leq$ -1) being a member of a given population given its metallicity and galactic space velocity~\citep{Grether_Lineweaver_2007}. The dashed and dotted lines indicate the uncertainty on the probabilities and are based on the uncertainty on the space velocities.}
\label{fig:thick}
\end{figure}

\noindent Following the method outlined by~\citet{Grether_Lineweaver_2007} that uses the upper limit on the metallicity for PG\,1018--047 (Table~\ref{tbl:petertable2}) together with the galactic space velocity, we can calculate the probability that the system is a member of the thin disk, thick disk or the halo. \citet{Grether_Lineweaver_2007} assume that all three populations are represented by a Gaussian distribution in the three space velocities and compare the observed space velocity and the metallicity (as defined by \citealt{Robin2003}) with the properties of the three populations. Plotting these probabilities for PG\,1018--047 in Figure~\ref{fig:thick}, membership of the thin disk can be excluded. Having only an upper limit on the metallicity available (see also Section~\ref{sec:peter}), however, we cannot differentiate between membership of the thick disk and the halo. Halo membership, though, is only possible in case PG\,1018--047 turns out to have a very low metallicity ([Fe/H] < -1.8).

\subsection{A post-EHB star?}

\begin{figure}
\centering
\includegraphics[width=\columnwidth]{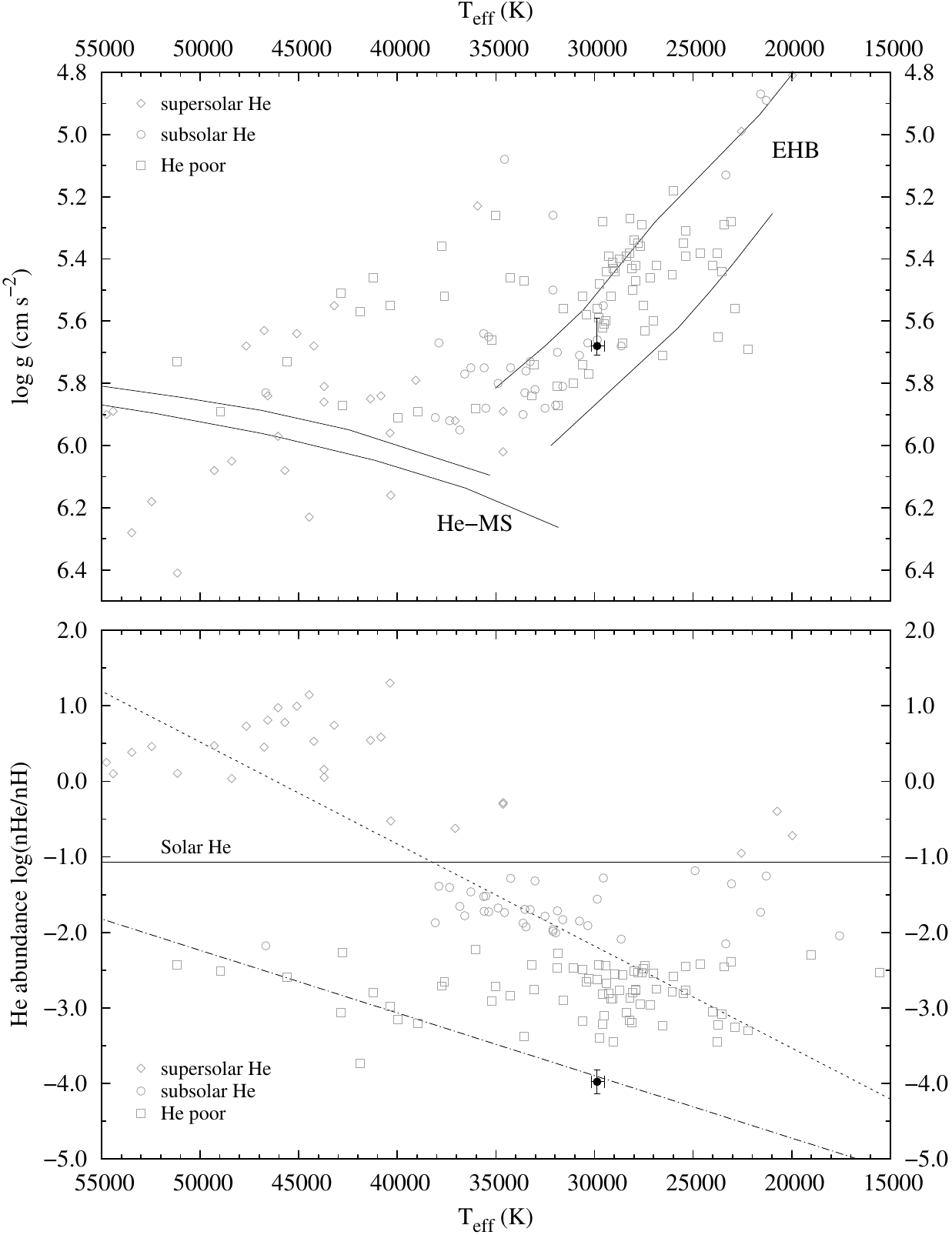}
\caption[]{The position of PG\,1018-047 in the $T_{\rm eff}-\log{g}$ and $T_{\rm eff}-\log({n{\rm He}/n{\rm H}})$ diagram among other subdwarfs from the $GALEX$ survey \citep{Nemeth_etal_2012}. 
The low He abundance of the sdB star places it onto the He-weak sequence. }
\label{fig-tghe}
\end{figure}

\noindent Could the sdB star be a low-mass white dwarf instead? The location of PG1018-047 in the $T_{\rm eff}-\log{g}$ diagram is consistent with an EHB star, but the surprisingly low He abundance is unusual, suggesting that the sdB star may have a different evolutionary history than the bulk of similar hot subdwarfs, or maybe it has already started evolving off the EHB? Note, also the atmospheric RLOF models of~\citet{Chen_etal_2013, Chen_etal_2014} predict a much higher metallicity ($Z\,=\,0.01-0.02$) for a binary system with a 752-day orbital period.\\

\noindent \citet{Edelmann_etal_2003} discovered that two distinct He sequences exist in the distribution of sdB stars in the $T_{\rm eff}-\log{n{\rm He}/n{\rm H}}$ abundance diagram, Figure \ref{fig-tghe}. In general, these parallel sequences show that hotter stars have higher He abundances and they are more compact. \citet{Nemeth_etal_2012} found that the He-rich sequence coincides with core He-burning EHB stars while the He-weak sequence represents post-EHB stars associated with a He shell-burning phase. There are only a few stars in between the two sequences and they show a large scatter towards low He abundances. Therefore the temperature and low He abundance of PG\,1018--047 may correspond to a post-EHB phase, although its surface gravity is more consistent with an EHB phase. \\

\noindent In the transitional stage between core and shell He-burning, stars are heating up ($\Delta{T}$\,$\sim$\,$10\,000$\,K), contracting ($\Delta\log{g}$\,$\sim$\,$0.4$\,dex) and the abundance trends suggest that their atmospheric He is lower. Atomic diffusion is the main process responsible for the peculiar abundance pattern of sdB stars \citep{Michaud_etal_2011}. If we assume that after He burning stops and the stellar luminosity drops the equilibrium between radiative levitation and gravitational settling breaks down, it could explain the observed low He abundance. In this context PG\,1018--047 could have evolved from the group of sdB stars near 28\,000 K and $\log{n{\rm He}/n{\rm H}}=-2.7\pm0.5$, and could then be associated with a post-EHB binary. \\

\section{Conclusions}~\label{sec5}

\noindent Based on 20 new high-resolution VLT/UVES spectra we have revisited the sdB+K5V-star long-period binary PG\,1018--047. We have refined the orbital period to $P\,=\,751.6\pm1.9\,\rm{d}$, in good agreement with~\citet{Deca_etal_2012}. In contrast to our earlier intermediate dispersion results, we determine that the binary is significantly eccentric ($e\,=\,$0.049$\pm0.008$). We accurately constrain the mass ratio of the system, $M_{\rm MS}/M_{\rm sdB} = 1.52 \pm 0.04$. The revised NLTE atmospheric parameters are $T_{\rm eff}$\,=\,29900$\pm$330\,K, log\,$g$ =\,5.65$\pm$0.06\,dex and log($n_{\rm He}$/$n_{\rm H}$)\,=\,--3.98$\pm$0.16\,dex.\\

\noindent With a companion mass of only $\sim$0.7$\rm\,M_{\odot}$, the system is at the lower companion-mass limit to support stable Roche-lobe overflow. Our MESA modelling analysis, given the observed eccentricity, shows that both phase-dependent mass loss during RLOF on its own, when 30 to 40\% of the available mass is lost during RLOF through the outer Lagrange point, and phase-dependent mass loss during RLOF in combination with a CB disk of maximum $M_{CB} = 0.001\rm\,M_{\odot}$ could have formed the PG\,1018--047 binary system.\\

\section*{Acknowledgments}
This work is based on observations collected at the European Organization for Astronomical Research in the Southern Hemisphere 
under ESO programmes 386.D-0224(A), 087.D-0231(A), 088.D-0041(A), 089.D-0921(A) and 090.D-0059(A).
JV acknowledges financial support from FONDECYT/CONICYT in the form of grant number 3160504.
This research has made use of the SVO Filter Profile Service supported from the Spanish MINECO through grant AyA2014-55216.
This research has used the services of {\sc Astroserver.org} under reference L0BDZP. \\












\bsp	
\label{lastpage}
\end{document}